\documentstyle[epsf,aps,prb]{revtex}

\newcommand{\be}{\begin{eqnarray}}
\newcommand{\en}{\end{eqnarray}}

\begin{document}
\bibliographystyle{prsty}
\draft

\title{Why are Orbital Currents Central to High T$_c$ Theory?}

\author{Patrick A. Lee and Guobin Sha}

\address{Center for Materials Science and Engineering and Department
of Physics,
MIT,
Cambridge, Massachusetts 02139}

\date{\today}
\maketitle

\begin{abstract}
We explain qualitatively why the staggered flux state plays a central
role in the SU(2) formulation of the $t$-$J$ model, which we use to
model
the pseudogap state in underdoped cuprates.  This point of view is
supported by studies of projected wavefunctions.  In addition to
staggered orbital current correlations, we present here for the first
time results of correlations involving hole and spin chirality and show
that the two are closely related.  The staggered flux state allows us
to
construct cheap and fast vortices, which may hold the key to explaining
the many anomalous properties of the normal state.

\end{abstract}

\vskip 0.5 in

\baselineskip15pt

It is now a widely accepted view that the problem of high T$_c$
superconductivity is the problem of doping into a Mott insulator.
By doping $x$ holes per unit cell, the N\'{e}el order is rapidly
destroyed and $d$-wave superconductivity emerges.  Many of us
believe the physics is captured by the $t$-$J$ model, and the
competition is between the kinetic energy of the hole $xt$, where
$t$ is the hopping matrix element, and the exchange energy $J$.
This competition leads to spectacular new physics in the
underdoped region, where the pseudogap phenomenon has been well
documented.  An understanding of the underdoped region is
prerequisite to understanding the entire phase diagram.

One view of the pseudogap phase is that it is a local
superconductor with robust amplitude but strong phase
fluctuations.  Setting aside the question of where the strong
pairing amplitude comes from in the first place, that this view is
incomplete can be seen from the following argument.  In two
dimensions the destruction of superconducting order is via the
Berezinskii-Kosterlitz-Thouless (BKT) theory of vortex unbinding.
Above T$_c$ the number of vortices proliferate and the normal
metallic state is reached only when the vortex density is so high
that the cores overlap.  (There is considerable latitude in
specifying the core radius, but this does not affect the
conclusion.)  At lower vortex density, transport properties will
resemble a superconductor in the flux flow regime.  In ordinary
superconductors, the BKT temperature is close to the mean field
temperature, and the core energy rapidly becomes small.  However,
in the present case, it is postulated that the mean field
temperature is high, so that a large core energy is expected.
Indeed, in a conventional core, the order parameter and energy gap
vanish with an energy cost of $\Delta_0^2/E_F$ per unit area.
Using a core radius of $\xi = V_F/\Delta_0$, the core energy of a
conventional superconductor is $E_F$.  In our case, we may replace
$E_F$ by $J$.  If this were the case, the proliferation of
vortices will not happen until a high temperature $\sim J$
independent of $x$ is reached.  Thus for the phase fluctuation
scenario to work, it is essential to have ``cheap'' vortices, with
energy cost of order T$_c$.  Then the essential problem is to
understand what the vortex core is made of.  In the past several
years, Wen and Lee have developed an SU(2) formulation of the
$t$-$J$ model,\cite{1} and the staggered flux state has emerged as
the natural candidate for the competing state which makes up the
vortex core.  Indeed, Lee and Wen have successfully constructed a
``cheap'' vortex state.\cite{2}  Other possibilities, such as a
N\'{e}el ordered state or spin density waves or
dimers,\cite{3,4,5,6} have been proposed.  In our view, the
staggered flux phase has an advantage over other possibilities in
that its excitation spectrum is similar to the $d$-wave
superconductor.  In any event the theory is not complete until the
nature of the alternative state which constitutes the vortex core
is understood.  Then the pseudogap phase can be understood equally
well as fluctuating superconductors with regions of the
alternative state or as a fluctuating alternative state with
regions of superconductivity.

The staggered flux state was first introduced as a mean field
solution at half-filling\cite{7}
and later was extended to include finite doping.\cite{8}  It exhibits
the remarkable property
that fermions hopping on a square lattice penetrated by staggered
flux $\pm \Phi_o$ has
an excitation spectrum identical to that of a $d$-wave superconductor
at half-filling.
There is a gap $\Delta_o$ at $(0,\pi)$ and nodes at $\left( \pm
{\pi\over 2},\pm
{\pi\over 2} \right)$, provided that $\tan \left( {\Phi_o \over
4}\right) = {\Delta_o \over
\chi_o } $, where $\chi_o$ is the hopping amplitude.  At
half-filling, due to the
constraint of no double occupation, the staggered flux state
corresponds to an insulating
state with power law decay in the spin correlation function.  It is
known that upon
including gauge fluctuations which enforce the constraint, the
phenomenon of confinement
and chiral symmetry breaking occurs, which directly corresponds to
N\'{e}el ordering.\cite{9}
The idea is that with doping, confinement is suppressed at some
intermediate energy
scale, and the state can be understood as fluctuating between the
staggered flux state
and the $d$-wave superconducting state.  Finally, when the holes
become phase coherent,
the $d$-wave superconducting state is the stable ground state.  Thus
the staggered flux
state may be regarded as the ``mother state'' which is an unstable
fixed point due to
gauge fluctuations.  It flows to N\'{e}el ordering at half-filling
and to the $d$-wave
superconductor for sufficiently large $x$.  Thus the staggered flux
state plays a central
role in this kind of theory.  We should point out that the staggered
flux state (called the $D$-density wave
state) has recently been proposed as the ordered state in the
pseudogap region.\cite{10}  As explained
elsewhere,\cite{11} we think that this view is not supported by
experiment and we continue to favor the
fluctuation picture.

The above picture finds support from studies of projected
wavefunctions, where the no-
double-occupation constraint is enforced by hand on a computer.  This
field has a long
history.\cite{12}  For example, it has been known for a long time
that if one takes a spin density wave state
for the fermion (i.e., introduce a staggered magnetization mean field
as a variational
parameter) and performs what is called the Gutzwiller projection,
i.e., project out all
doubly-occupied configurations, one does not obtain a very good
wavefunction.  On the
other hand, projection of a $\pi$-flux phase without any variational
parameter does surprisingly well.
   The best state is the combination of staggered magnetization with
some flux, either
$\pi$-flux or staggered-flux, and an excellent energy is achieved.
With doping the best state
is a projected
$d$-wave state. \cite{12}  Not surprisingly, this state has long
range pairing order after
projection.  Recently we calculated the current-current correlation
function of this state
\cite{13}

\begin{equation}
c_j(k,\ell) = < j(k)j(\ell) >
\end{equation}
where $j(k)$ is the physical electron current on the bond $k$.  The
average current $<j(k)>$ is
obviously zero, but the correlator exhibits a staggered circulating
pattern as shown in Fig.1.
(Note that the pattern is shifted by $\pi$ relative to a pattern
constructed from the reference
bond at the origin.)  Within our numerical accuracy, this correlation
decays as a power law and the
decay is faster with increasing doping.  Such a pattern is absent in
the $d$-wave BCS state before
projection, and is a result of the Gutzwiller projection.

\begin{figure}[h]
\epsfxsize=2.25truein
\centerline{\epsffile{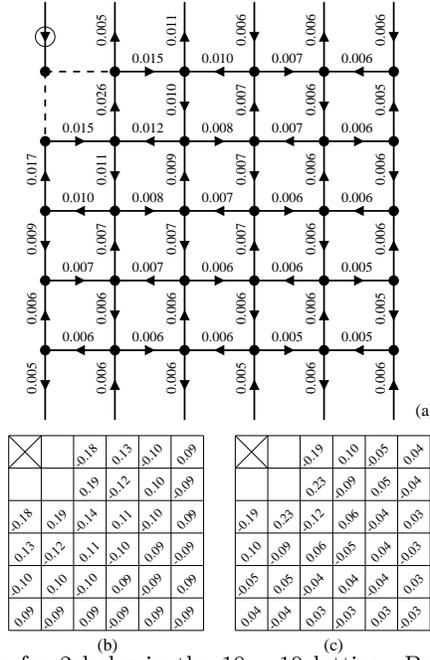}}
\caption{(a) Current-current correlations for 2 holes in the $10
\times 10$ lattice.  Boundary conditions are periodic in one and
antiperiodic in the other directions (the data are averaged over
the two orientations).  The number on a link is the correlation of
the current on this link and of the current on the circled link
divided by hole density.  The arrows point in the direction of the
positive correlations of the current.  (b) The same data in the
form of vorticity defined as the sum of the current around a
plaquette.  The number on a plaquette is the vorticity correlation
divided by $x$ with the crossed plaquette.  (c) Same as (b) for 10
holes in $10 \times 10$ lattice.}
\end{figure}

We were motivated to look for the staggered pattern in the
current-current correlation function
because that is what we expect to find in the staggered flux phase.
Consider a plaquette with a
hole in the top left corner (4) and spins on the other three corners
(labelled clockwise 1--3).  A hole
hopping around the plaquette sees a wandering spin quantization axis
from site to site and will pick up a
Berry's phase $\Phi$ which is given by the solid angle subtended by
the 3 spin directions.
\cite{14}  The flux in the flux phases is designed to capture this
piece of physics, as a hole
hopping in the presence of a gauge ``magnetic flux'' will also pick
up an Aharonov-Bohm phase.
The physical idea is that the N\'{e}el state is detrimental to hole
hopping, but if the spins are not
coplanar, we may achieve a better compromise between the exchange
energy and the hole kinetic energy. More
formally, in the slave boson representation where the electron
operator is written as $c_{i\sigma} =
f_{i\sigma}b_i^\dagger$, it is known
\cite{14,15} that if we define
$
Q_{ij} = \sum_\sigma f_{i\sigma}^\dagger f_{j\sigma}
$,
then

\begin{eqnarray}
Im \left(
\prod_\Box : Q_{12} Q_{23} Q_{34} Q_{41} :
\right)
& = &
\left(
f_{4\sigma}^\dagger f_{4\sigma}
\right)
\vec{S}_1 \cdot \vec{S}_2 \times \vec{S}_3 \\ \nonumber
& & + \,\, \mbox{{\rm permutations}} \,\,\,\, .
\end{eqnarray}
In the mean field theory, the L.H.S. of Eq.(2) is $ |\chi_o |^4
\sin \Phi $. Thus the flux is also related to the spin chirality
defined as $\chi = \vec{S}_1 \cdot \vec{S}_2 \times \vec{S}_3$. We
note that in Eq.(2), $ f_{4\sigma}^\dagger f_{4\sigma} = 1 -
b^\dagger b = 1 - n_H $ where $n_H$ is the hole density.  We thus
consider

\begin{equation}
c_{\chi_H \chi_H}  (k,\ell) = \left< n_{H} (4) \vec{S}_1 \cdot
\vec{S}_2 \times \vec{S}_3 (k) \,\,\, n_{H}(4^\prime)
\vec{S}_{1^\prime} \cdot \vec{S}_{2^\prime} \times
\vec{S}_{3^\prime}(\ell) \right> \,\,\,\, .
\end{equation}
where $(k,\ell)$ labels plaquettes and the spins 1,2,3 form a
triangle with a fixed orientation around plaquette $k$, as
described earlier.

\begin{figure}[h]
\epsfxsize=400pt \centerline{\epsffile{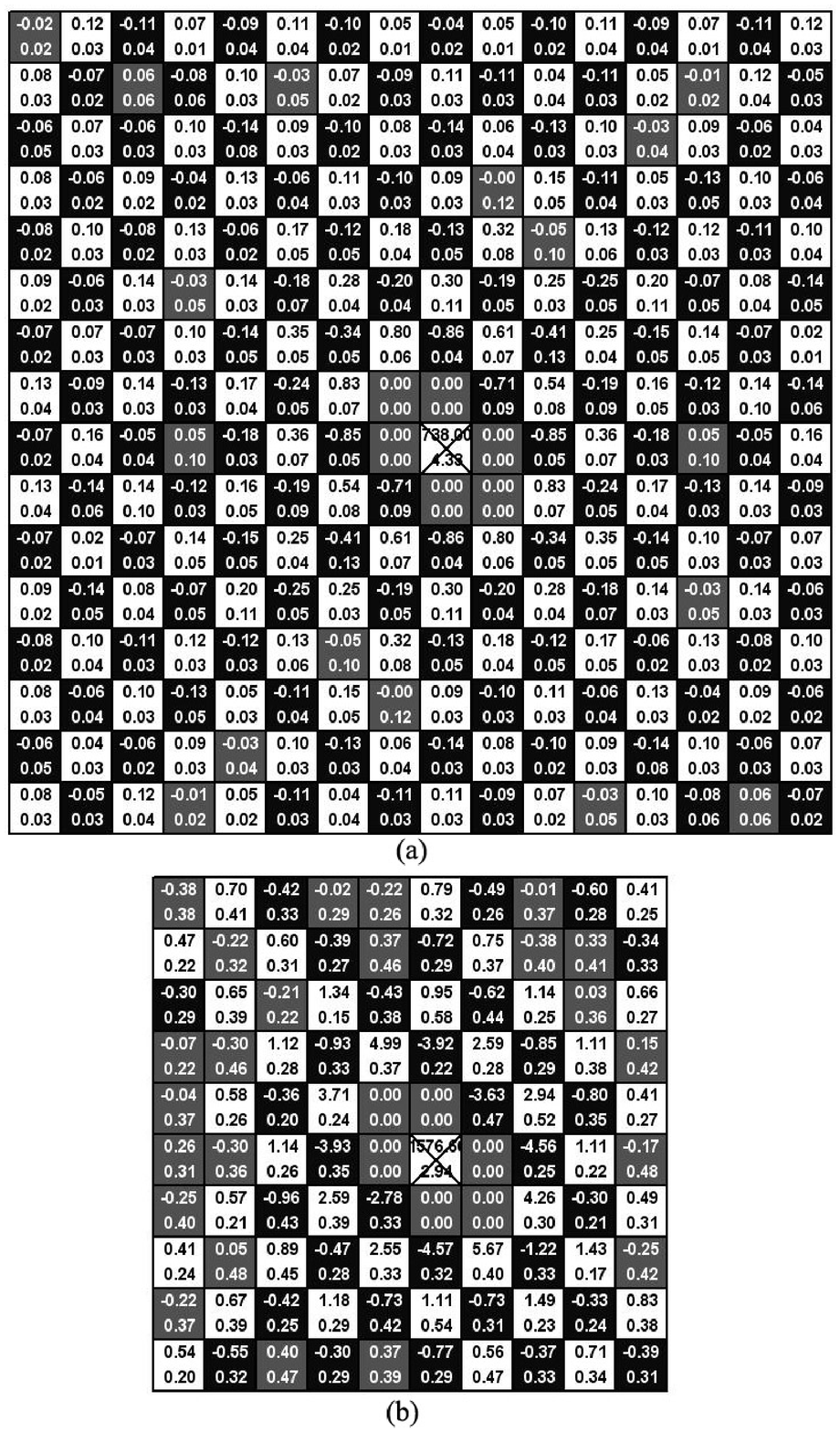}}

Fig.2.  The correlation function $c_{\chi_H \chi_H} \times {1\over
x} $ as a function of plaquette position. The reference plaquette
is at the center. The lower number is the statistical error.
Plaquettes with positive (negative) values are shown in white
(black) while plaquettes whose signs are undetermined within the
error bar are shown in grey.   The calculation was done with
periodic boundary conditions in one direction and antiperiodic
boundary conditions in the other.  The only symmetry is between
$\bbox{r}$ and $\bbox{-r}$ and this symmetry was used in the
computation. Fig.2(a) is for two holes in a $16 \times 16$ lattice
and Fig.2(b) is for six holes in a $10 \times 10$ lattice. The
correlation data is in unit of $10^{-5}$.
\end{figure}

 In Fig.2 we show the results for $c_{\chi_H \chi_H}$
for two holes in $16 \times 16$ and six holes in $10 \times 10$.
The large number  of operators in Eq.(3) makes the computation
more time consuming than for $c_j$ and the resulting error bars
are larger.   Setting aside results with undetermined signs within
the error bar, we find a perfect staggered correlation given by
the black and white checkerboard pattern. The pattern is also
phase shifted from the central one, just as we found for $c_j$.
Note that the correlator on the same site is very large. This is
because it only require the presence of a hole on a single site.
To get a fair comparison, we should multiply the equal site
correlator by a factor of $x$. Note that the correlator decreases
rather slowly with distance. This decrease is shown in Fig.3,
which plots $c_{\chi_H \chi_H }$ vs $R^2$ on a log-log plot. We
also plot the vorticity correlator constructed from $c_j$ from
ref. (13). We see that both are consistent with the same power law
decay.\cite{13}

\begin{figure}[h]
\epsfxsize=400pt \centerline{\epsffile{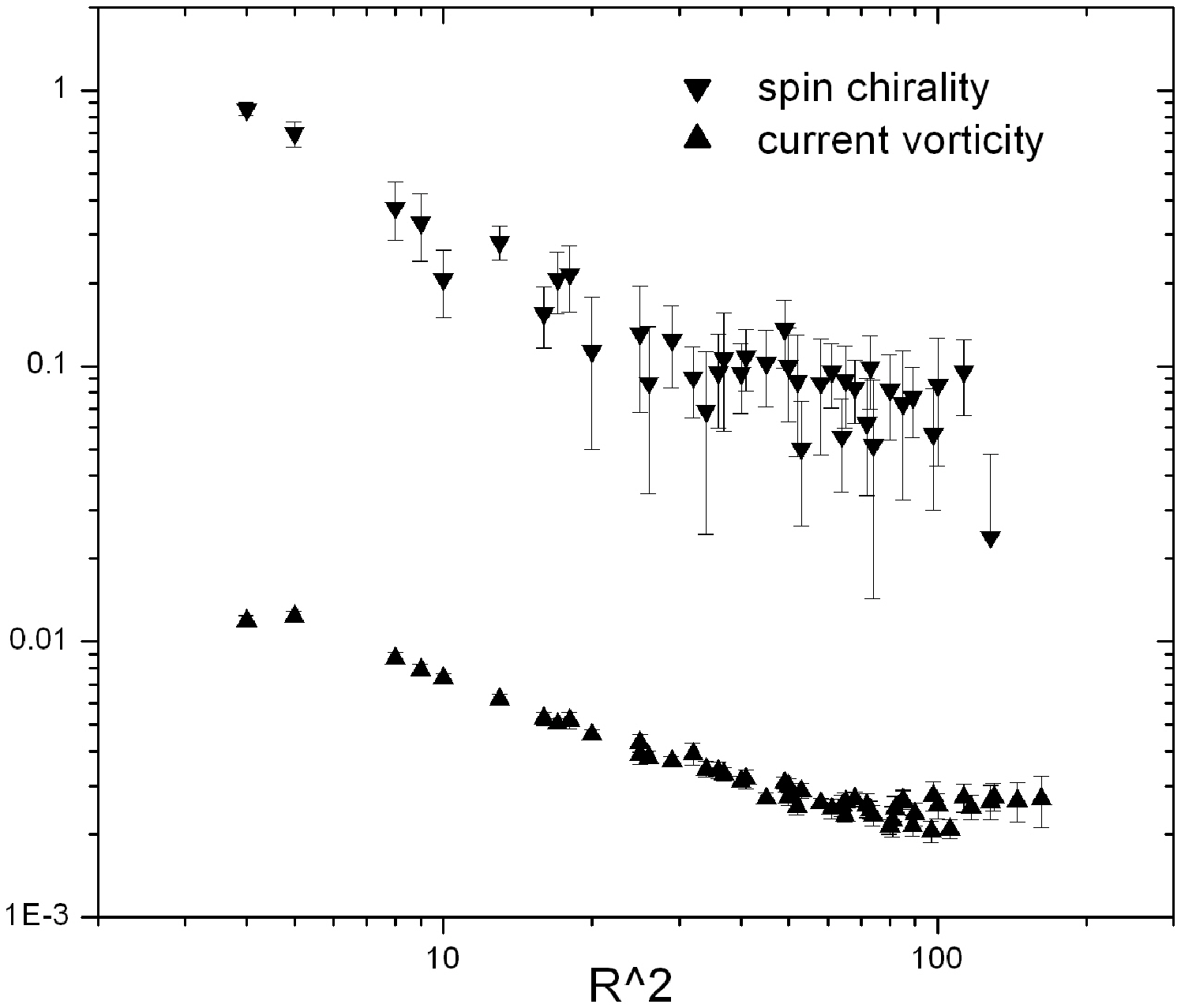}}
Fig.3  The absolute value of the correlation function $c_{\chi_H
\chi_H} \times {1\over x} $ plotted vs the square of the distance
between plaquettes on a log-log plot for two holes in a $16 \times
16$ lattice. Values of plaquettes with the same distance have been
averaged. The correlation data is in unit of $10^{-5}$. We also
reproduced the data on the current vorticity correlator divided by
$x$ for 2 holes in $18\times 18$ sites from Ivanov et al.
\cite{13} We have divided the latter by 10 for clarity.
\end{figure}

We have also computed $c_{\chi_H \chi}$ given by
\begin{equation}
c_{\chi_{H} \chi} (k,\ell) = \left< n_{H} (4) \vec{S}_1 \cdot
\vec{S}_2 \times \vec{S}_3 (k) \,\,\, \vec{S}_{1^\prime} \cdot
\vec{S}_{2^\prime} \times \vec{S}_{3^\prime}(\ell) \right>
\,\,\,\, .
\end{equation}
This gives information about the spin chirality generated by a
hole and chirality combination some distance away. Naively we
might expect that $c_{\chi_H \chi}$ should be larger than
$c_{\chi_H \chi_H}$ by a factor ${1\over x}$, because a hole is
not required to be associated with one of the chirality factors.
Surprisingly we found that $c_{\chi_H \chi}$ is of similar
magnitude as $c_{\chi_H \chi_H}$ and shows an intricate
correlation. However, due to the small size of the correlation
that results from delicate cancellations, the error bars are too
large for us to make a detailed analysis.  Our results for $c_j$,
$c_{\chi_H \chi_H}$, and $c_{\chi_H \chi}$ are consistent with
exact diagonalization of two holes in 32 sites.\cite{16}

We conclude that the correlation of the hole with the spin
chirality on the same plaquette is the key physics and that the
current correlator may be viewed as the symptom, rather than the
root cause of the pseudogap phenomenon.  Of course, the current
correlator has the advantage that it is easier (relatively
speaking) to measure than the $\chi_H \chi_H$ correlator.  The
staggered current generates a staggered physical magnetic field
(estimated to be 10--40 gauss)\cite{8,13} which may be detected,
in principle, by neutron scattering.  In practice the small signal
makes this a difficult, though not impossible experiment and we
are motivated to look for situations where the orbital current may
become static or quasi-static.  Recently, we analyzed the
structure of the $hc/2e$ vortex in the superconducting state
within the SU(2) theory and concluded that in the vicinity of the
vortex core, the orbital current becomes quasi-static, with a time
scale determined by the tunnelling between two degenerate
staggered flux states.\cite{2} It is very likely that this time is
long on the neutron time scale.   Thus we propose that a
quasi-static peak at  $(\pi,\pi)$ will appear in neutron
scattering in a magnetic field, with intensity proportional to the
number of vortices.  The time scale may actually be long enough
for the small magnetic fields generated by the orbital currents to
be detectable by $\mu$-SR or Yttrium NMR.  Again, the signal
should be proportional to the external fields.  (The NMR
experiment must be carried out in 2--4--7 or 3 layer samples to
avoid the cancellation between bi-layers.)  We have also computed
the tunnelling density of states in the vicinity of the vortex
core, and predicted a rather specific kind of period doubling
which should be detectable by atomic resolution STM.\cite{17}  The
recent report\cite{18} of a static field of $\pm 18$ gauss in
underdoped YBCO which appears in the vortex state is promising,
even though muon cannot distinguish between orbital current or
spin as the origin of the magnetic field.  We remark that in the
underdoped antiferromagnet, the local moment gives rise to a field
of 340 gauss at the muon site.  Thus if the 18 gauss signal is due
to spin, it will correspond to roughly $1/20$th of the full
moment.

We remark that our analytic model of the vortex core is in full
agreement with the numerical solution of
unrestricted mean field $\Delta_{ij}$ and $\chi_{ij}$ by Han and
Lee.\cite{4}  Upon re-examination of their
numerical solution, they also found staggered orbital current in
their vortex core.\cite{19}  This vortex
solution is also interesting in that the tunnelling density of states
show a gap, with no sign of the large
resonance associated with Caroli-deGennes-type core levels found in
the standard BCS model of the
vortex.\cite{20}  There exists a single bound state at low lying
energy,\cite{19,21} in agreement with STM
experiments.\cite{22}  The low density of states inside the vortex
core has an important implication.  In
the standard Bardeen-Stephen model of flux-flow resistivity, the
friction coefficient of a moving vortex is
due to dissipation associated with the vortex core states.  Now that
the core states are absent, we can
expect anomalously small friction coefficients for underdoped
cuprates.  The vortex moves fast transverse to
the current and gives rise to large flux-flow resistivity.  Indeed,
the flux-flow resistivity is given by

\begin{displaymath}
\rho_{\bbox{\rm{flux-flow}} } = {B \Phi_o \over\eta c^2} \,\,\, .
\end{displaymath}
Since the total conductivity is the sum of the flux-flow conductivity
and the quasiparticle conductivity, it
is possible to get into a situation where the quasiparticle
conductivity dominates even for H $\ll$
H$_{c2}$.   Thus the ``cheap'' and ``fast'' vortex opens the
possibility of having vortex states above the
nominal T$_c$ and H$_{c2}$, when the resistivity looks like that of a
metal, with little sign of flux-flow
contribution.\cite{23}  From this point of view, the large Nerst
effect observed by Ong and
co-workers\cite{24} over a large region in the H-T plane above the
nominal T$_c$ and H$_{c2}$ (as determined
by resistivity) may be qualitatively explained.

\vskip 0.25 in
We thank X.-G. Wen and D. Ivanov for collaborating on much of the
work reported here.  We acknowledge the
support of NSF through the MRSEC program under grant No. DMR98-08941.

\end{document}